\documentstyle[preprint,aps,floats,psfig]{revtex}
\catcode`@=11
\def\references{%
\ifpreprintsty
\bigskip\bigskip
\hbox to\hsize{\hss\large \refname\hss}%
\else
\vskip24pt
\hrule width\hsize\relax
\vskip 1.6cm
\fi
\list{\@biblabel{\arabic{enumiv}}}%
{\labelwidth\WidestRefLabelThusFar  \labelsep4pt %
\leftmargin\labelwidth %
\advance\leftmargin\labelsep %
\ifdim\baselinestretch pt>1 pt %
\parsep  4pt\relax %
\else %
\parsep  0pt\relax %
\fi
\itemsep\parsep %
\usecounter{enumiv}%
\let\p@enumiv\@empty
\def\theenumiv{\arabic{enumiv}}%
}%
\let\newblock\relax %
\sloppy\clubpenalty4000\widowpenalty4000
\sfcode`\.=1000\relax
\ifpreprintsty\else\small\fi
}
\catcode`@=12
\begin{document}
\def\mh{m_h^{}}
\def\vev#1{{\langle#1\rangle}}
\def\gev{{\rm GeV}}
\def\tev{{\rm TeV}}
\def\fbi{\rm fb^{-1}}
\def\lsim{\mathrel{\raise.3ex\hbox{$<$\kern-.75em\lower1ex\hbox{$\sim$}}}}
\def\gsim{\mathrel{\r6aise.3ex\hbox{$>$\kern-.75em\lower1ex\hbox{$\sim$}}}}
\newcommand{\teb}{{T_{\bar{\nu}_e}}}
\newcommand{\neb}{{\bar{\nu}_e}}
\newcommand{\tneb}{{T_{\bar{\nu}_e}}}
\newcommand{\tnx}{{T_{\bar{\nu}_x}}}
\newcommand{\eneb}{{\vev{E_{\bar{\nu}_e}}}}

\newcommand{ \slashchar }[1]{\setbox0=\hbox{$#1$}   
   \dimen0=\wd0                                     
   \setbox1=\hbox{/} \dimen1=\wd1                   
   \ifdim\dimen0>\dimen1                            
      \rlap{\hbox to \dimen0{\hfil/\hfil}}          
      #1                                            
   \else                                            
      \rlap{\hbox to \dimen1{\hfil$#1$\hfil}}       
      /                                             
   \fi}                                             %

\tighten
\preprint{ \vbox{
\hbox{MADPH--02-1265}
\hbox{AMES-HET-02-03}
\hbox{hep-ph/0204253}}}
\title{Imprint of SNO neutral current data on the\\ solar neutrino problem}
\author{$^1$V. Barger, $^2$D. Marfatia, $^3$K. Whisnant and $^1$B. P. Wood}
\vskip 0.3in
\address{$^1$Department of Physics, University of Wisconsin, Madison, WI 53706}
\vskip 0.1in
\address{$^2$Department of Physics, Boston University, Boston, MA 02215}
\vskip 0.1in
\address{$^3$Department of Physics and Astronomy, Iowa State University, 
Ames, IA 50011}
\maketitle

\begin{abstract}
{\rm We perform a global analysis in the framework of two 
active neutrino oscillations  of all solar neutrino data,
including the recent SNO day and night spectra 
(comprised of the charged current (CC), elastic scattering (ES) and 
 neutral current (NC) events), the Super-Kamiokande
(SK) day and night spectra (from 1496 days) and the updated SAGE results. 
We find that the Large 
Mixing Angle (LMA) solution is selected at the 99\% C.L.; 
the best-fit parameters 
are $\Delta m^2=5.6 \times 10^{-5}$ eV$^2$ and 
$\theta=32^{\circ}$. 
No solutions with $\theta\geq \pi/4$ are allowed at the 5$\sigma$ C.L.
Oscillations to a pure sterile state are
excluded at 5.3$\sigma$, but a sizeable sterile neutrino component could
still be present in the solar flux.
}
\end{abstract}

The end of the era of the solar neutrino problem is on the horizon. Neutrino oscillations
are a compelling explanation of the solar neutrino deficit relative to the Standard Solar
Model~\cite{SSM} found by several experiments~\cite{cl,gallex,gno,sage,sk,sno,sno1}. The
SNO experiment has provided cogent evidence in favor of this hypothesis by separately
measuring the incident $\nu_e$ flux via the CC reaction and the total active neutrino flux 
via the NC reaction in the same energy range~\cite{sno}. With the continued
 accumulation of 
solar neutrino data, the LMA solution has become the 
most favored. If LMA is in fact the solution, 
the ongoing KamLAND experiment~\cite{kamland} will
provide a precise and SSM-independent measurement of the oscillation parameters by measuring
the suppression and distortions of the $\bar{\nu}_e$ flux emerging from the surrounding
nuclear reactors~\cite{bmw}. 
Aside from a possible complication of partial oscillations to sterile neutrinos,
the thirty year old solar neutrino problem will be solved.

Here we assess the extent to which the recent SNO results home in on an
 unique oscillation
solution by performing a global analysis of all solar neutrino data in a 
two neutrino
framework. There is little difference between a three neutrino analysis and
an effective two neutrino analysis because the mixing between the solar
and atmospheric scales is known to be small~\cite{chooz}.
We first show through a
model-independent analysis that pure active to sterile oscillations are excluded at 
high confidence and then proceed with a global analysis  
involving only active flavors.

A universal sum rule that
 holds for arbitrary active-sterile oscillation admixtures 
has been derived~\cite{lma} which relates the NC flux at SNO
to the CC flux at SNO and the neutrino flux measured at SK with elastic scattering:
\begin{equation}
\Phi_{NC}=[\Phi_{SK}-(1-r)\Phi_{CC}]/r \,,
\end{equation}
where $r\equiv \sigma_{\nu_\mu,\nu_\tau}/\sigma_{\nu_e}$ is the ratio of the 
$\nu_{\mu,\tau}$ to ${\nu_e}$ elastic scattering cross sections on electrons.
Above the 5 MeV threshold of the two experiments, $r=1/6.48$ and the relation becomes
\begin{equation}
\Phi_{NC}=6.48\Phi_{SK}-5.48\Phi_{CC} \,.
\label{sum}
\end{equation}
The values from the data (assuming an undistorted $^8$B 
spectrum~\cite{sno,sno1,sno2}), 
for the left and
right-hand sides of Eq.~(\ref{sum}) are, respectively,
\begin{equation}
5.09 \pm 0.62 \ \ \ \ \ \ \ \,, \ \ \ \ \ \ \ \ 5.58 \pm 0.71
\end{equation}
showing agreement between the data 
and the sum rule. The fact that the two values have
similar percentage uncertainties ($\sim$ 12-13\%) shows that the accuracy of the SNO NC measurement is already
comparable to that which can be inferred from the SNO CC and SK data.
 
The SNO NC rate is a measure of the flux of active neutrinos in the high energy part
of the solar neutrino spectrum. If an active-sterile admixture is responsible for
the solar neutrino deficit,
\begin{equation}
{\Phi_{NC} \over\Phi_{SSM}}=\beta {\Phi_{SSM}-\Phi_{\nu_s} \over \Phi_{SSM}}\,,
\end{equation}
with measured value (see Table~\ref{tab1}),
\begin{equation}
{\Phi_{NC} \over\Phi_{SSM}}=1.01\pm 0.12 \,.
\end{equation}
\begin{table}[ht]
\caption{Solar neutrino data relevant to our analysis. Note that we use the 
SNO day and night spectra in place of the SNO CC, NC and $A_{DN}$ values and 
SK day and night
spectra instead of the SK rate. 
The SNO CC, NC and $A_{DN}$ numbers were extracted from the SNO day and 
night spectra assuming an undistorted $^8$B spectrum and are given here only
for reference as is the value of $\Phi_{SSM}$. 
1 SNU is 1 interaction/s$/10^{36}$ atoms of the
neutrino absorbing isotope. 
 }
\begin{center}
\begin{tabular}{|ll|} 
\hline 
          & Measurement\\ \hline
Homestake &$2.56 \pm 0.23$ SNU \\ 
GALLEX+GNO&$73.3 \pm 4.7 \pm 4.0$ SNU \\
SAGE      &$70.8 \pm 5.3 \pm 3.5$ SNU   \\  
SK        &$2.35 \pm 0.02 \pm 0.06 \times 10^6$ cm$^{-2}$ s$^{-1}$ \\
SNO CC    &$1.76 \pm 0.06 \pm 0.09 \times 10^6$ cm$^{-2}$ s$^{-1}$ \\
SNO NC    &$5.09 \pm 0.44 \pm 0.45 \times 10^6$ cm$^{-2}$ s$^{-1}$ \\
SNO $A_{DN}$&$0.07 \pm 0.049 \pm 0.013$\\
$\Phi_{SSM}$&$5.05(1 \pm 0.18)\times 10^6$ cm$^{-2}$ s$^{-1}$
\end{tabular}
\end{center}
\label{tab1}
\end{table}  
Here $\beta$ is a normalization of the $^8$B flux with respect to the central 
value of the SSM 
prediction $\Phi_{SSM}=5.05\times 10^6$ cm$^{-2}$s$^{-1}$ 
and $\beta \Phi_{\nu_s}$ is the total sterile neutrino flux to which 
the electron neutrinos oscillate{\footnote{As in the SNO analysis~\cite{sno} 
we conservatively adopt the SSM $^8$B flux of BPB2000~\cite{SSM} since the
$S_{17}(0)$ determination~\cite{s17} 
used in Ref.~\cite{robust} is being reanalyzed~\cite{int}.}.
Since the measured 
$\Phi_{NC}$ is consistent with the
SSM prediction, 
we impose the SSM constraint $\beta=1\pm 0.18$, and obtain
\begin{equation}
{\Phi_{\nu_s} \over \Phi_{SSM}}= P(\nu_e \rightarrow \nu_s)= -0.01 \pm 0.22\,.
\end{equation}
While the central value of the SNO NC rate suggests a solution
with oscillations only to active flavors, the uncertainty is too large 
to rule out a substantial
sterile neutrino flux on this basis alone{\footnote{Assuming an LMA solution
and a small sterile admixture, a combination of KamLAND and SNO CC data can
determine $\beta$ to 11\%~\cite{john}. This uncertainty is still too large 
to eliminate the possibility of 
a significant sterile component in the solar flux.}.

To see this freedom from another angle, we return to the approach of Ref.~\cite{lma}.
 Of the neutrinos that oscillate, the fraction that oscillate
 to active neutrinos is
\begin{equation}
\sin^2 \alpha={\Phi_{NC}-\Phi_{CC} \over \beta \Phi_{SSM}-\Phi_{CC}}=1.01 \pm 0.34\,.
\label{ster}
\end{equation}
Thus, the evidence for transitions to active flavors is at the 3$\sigma$ C.L.
However, large sterile fractions
 are allowed even at the 2$\sigma$  C.L. See Fig.~\ref{sterile} for 
illustration. The dark-shaded and light-shaded 
regions enclose the values of $\Phi_{NC}$ and $\Phi_{CC}$ allowed by the
SSM at 1$\sigma$ for $\sin^2 \alpha=1$ and $\sin^2 \alpha=0.5$, respectively. The lines
through the center of these bands correspond to the central value of the SSM $^8$B flux
prediction. 
The region above the diagonal, ${\Phi_{NC}=\Phi_{CC}}$, is forbidden
because $\Phi_{CC}>\Phi_{NC}$ is impossible. 
 The measured SNO CC and NC rates are marked by
a cross. Even a doubling of the widths of the SSM bands and the SNO rate 
uncertainties (effectively 2$\sigma$) allows large sterile fractions.

We emphasize that a large sterile neutrino flux is viable not
only for an analysis of the total rates, but also when the SK day and
night spectra are included (contrary to the assertion in Ref.~\cite{err}).
 For the issue in question, the effect of imposing 
the SSM $^8$B flux constraint is equivalent to including the day and night 
spectra: large
values of $\beta$ are not allowed and the best-fit value of $\sin^2 \alpha$ 
is close to pure active oscillations.  That this is the case can be seen from
Fig.~4 of Ref.~\cite{lma} ($^8$B flux constraint applied) 
and Fig.~2 of Ref.~\cite{john} (day and night spectra used). 
 The reason for this correlation is that the day and night
spectra rule out a large day-night effect, which is the same region of
the LMA solution (low $\Delta m^2$) that favors large $\beta$ for small $\sin^2 \alpha$.
 Therefore
in our rate analysis, imposing the $^8$B
flux constraint should yield the same effect as including the SK day and
night spectra, and we see from Eq.~(\ref{ster}) that a large sterile flux
is still allowed.

In the above analysis, pure active
and pure sterile oscillation solutions are treated on an equal footing.  
If instead we use the a priori $\beta$-independent criterion  
that for a pure sterile oscillation solution,
\begin{equation} 
\Phi_{NC}=\Phi_{CC}
\end{equation} 
(the diagonal of Fig.~\ref{sterile}), 
then such a solution is allowed only at the 5.3$\sigma$ C.L. 
In light of the strong evidence from SNO that oscillations to a pure sterile
state are not responsible for the solar anomaly, we hereafter 
only consider oscillations 
between active neutrinos. However, oscillations to an active-sterile
admixture remains viable and are worthy of investigation.

For a pure active oscillation solution, 
the NC rate measurement at SNO provides a 
direct determination of the $^8$B flux produced in the Sun
{\footnote{However, the NC rate derived from SNO's day and 
night spectra is dependent on the shape of the CC energy spectrum, and hence 
on the oscillation parameters~\cite{sno2}. As a result, the CC and NC fluxes 
are strongly statistically correlated}. This frees us from relying on the
Standard Solar Model (SSM) prediction of this flux which has a large 
uncertainty.
That is particularly significant because both SK and SNO 
are mainly sensitive to $^8$B neutrinos.
These experiments are also sensitive to $hep$ neutrinos, 
which according to the SSM constitute a very tiny fraction compared to 
the $^8$B neutrinos. 
According to Ref.~\cite{hep}, the
$hep$ flux is in fact the same as that of the SSM with an uncertainty of 
20\%. 
To avoid any dependence on the SSM $^8$B flux,
previous authors have performed $^8$B flux free analyses so as to extract the oscillation parameters
directly from the data~\cite{ind}. 
With the recent SNO results, this is no longer necessary; the NC
rate is itself a $^8$B flux measurement and thus a 
probe of the oscillation dynamics. We take this approach. Also, we fix the 
unoscillated $hep$ flux at the SSM value.

Global analyses of solar data are by now quite standard other than the $\chi^2$ function
employed for the statistical treatment~\cite{global}. We briefly describe the salient  
features of our analysis. We work in the framework
of oscillations between two active neutrino flavors and focus on the 
Mikheyev-Smirnov-Wolfenstein (MSW) solutions since the vacuum
solution is tenuous at best; for completeness we also analyze the data in the limit of a
pure vacuum solution. 
To evaluate the survival probability of solar neutrinos,
we consider neutrino production points that are nonradial and consider 
the possibility of
double resonances. The production point region of the different neutrino fluxes is as given 
by the SSM.  
We use semi-analytic expressions of the survival probability 
that have been derived for the almost exponential matter density of the 
Sun~\cite{analytic}, but numerically
integrate the evolution equations for the passage of neutrinos through the earth. For the
earth-matter density, we make use of the Preliminary Reference Earth Model~\cite{prem}. 
The time
spent by a detector at a particular zenith angle is given by the exposure 
function~\cite{exp}, which
determines the extent to which earth matter affects the survival probability.

For the unoscillated neutrino fluxes other than the $^8$B flux,
we adopt the SSM predictions from the $pp$ chain and $CNO$ cycle. We treat the
$^8$B flux normalization $\beta$ 
as a parameter that is constrained by the SNO NC measurement. The
undistorted spectrum shape of the $^8$B neutrinos is given in Ref.~\cite{B8}. 
To determine the
expected signal at each detector, the fluxes are convoluted with the survival probability
at the detector, the neutrino cross-sections and the detector response functions (for 
SK and SNO). We use the neutrino-electron elastic scattering cross-sections of 
Ref.~\cite{es}, and
the CC and NC cross-sections of neutrinos on deuterium of Ref.~\cite{cs}. The response functions
are given in Ref.~\cite{respsk} for SK and in Ref.~\cite{sno2} 
for SNO.

We analyze the event rate from the Homestake experiment~\cite{cl}, 
the combined rate from GALLEX and GNO~\cite{gallex,gno}, the latest 
SAGE event rate~\cite{sage}, the SK day and night spectra
corresponding to 1496 days of running~\cite{sk},
and the SNO day and night spectra 
(which include the CC, ES and NC fluxes)~\cite{sno,sno1,sno2}.
We do
not use the SK rate since the SK day and 
night spectra already include the normalization information~\cite{ind}. 

The statistical significance of an oscillation solution is determined by evaluating a 
suitably chosen $\chi^2$ function. We define a $\chi^2$
 that depends sensitively on the SNO NC flux
and and is completely independent of the 
SSM $^8$B prediction. It is
\begin{eqnarray}
\chi^2 &=&\sum_{i,j=1,3} (R_i^{th}(\beta)-R_i^{exp})
(\sigma_R^2)^{-1}_{i j}(R_j^{th}(\beta)-R_j^{exp})\nonumber \\ &&
+\sum_{i,j=1,38} (R_i^{th}(\beta)-R_i^{exp})(\sigma_{SK}^2)^{-1}_{i j}(R_j^{th}(\beta)-R_j^{exp})\nonumber\\ && + 
\sum_{i,j=1,34} (R_i^{th}(\beta)-R_i^{exp})(\sigma_{SNO}^2)^{-1}_{i j}(R_j^{th}(\beta)-R_j^{exp})\,.
\label{base}
\end{eqnarray}
In Eq.~(\ref{base}), 
$R_i^{th}$ and $R_i^{exp}$ denote the theoretical and experimental value of the 
event rate or flux measurement (depending on whether $i$ is an experiment or a spectrum 
bin) normalized to the expectation for no oscillations. 
The first term in $\chi^2$ is the 
contribution of the rate measurements to the analysis. The sum  
runs from 1 to 3 because
the Homestake, GALLEX+GNO and SAGE rates are included. The $3\times 3$ 
matrix $\sigma_R^2$ contains the experimental 
(statistical and systematic) and theoretical uncertainties. 
This matrix involves 
strong correlations arising from solar model parameters~\cite{corr}. 
Note that $\sigma_R^2$ does not include a theoretical
uncertainty for the $^8$B neutrino flux.

\begin{table}[ht]
\caption{Best fit values of $\Delta m^2$, $\tan^2 \theta$ and the $^8$B normalization 
$\beta$ for each solution from a global analysis. 
The corresponding NC/CC and $\chi^2$ values are tabulated 
in the last two columns. The
number of degrees of freedom is 72.}
\begin{center}
\begin{tabular}{|l|c|c|c|c|c|} 
\hline 
Solution & $\Delta m^2$ (eV$^2$) & $\tan^2 \theta$& $\beta$ & NC/CC& $\chi^2_{\rm{min}}$\\ \hline
LMA&$5.6\times 10^{-5}$ & 0.39   & 1.09 & 3.19 &50.7 \\ \hline
LOW &$1.1 \times 10^{-7}$ & 0.46 & 1.03 & 2.92 &59.9 \\ \hline
SMA &$7.9 \times 10^{-6}$ & $2.0\times 10^{-3}$ & 1.46 & 4.85 &   108\\ \hline
VAC &$1.6 \times 10^{-10}$ &  0.25 (3.98) & 0.89 & 2.46 &  76.3
\end{tabular}
\end{center}
\label{tab2}
\end{table}

The second term in Eq.~(\ref{base}) is the 
contribution of the distortions of the SK day and night spectra and 
implicitly, that of the SK rate.
$\sigma_{SK}^2$ is a $38\times 38$ matrix that contains
the statistical and systematic uncertainties of the $19+19$ spectral bins. The systematic 
uncertainties include the energy correlated, energy uncorrelated and energy independent 
contributions. 

The third term encapsulates the contributions of the SNO CC, ES and NC rates 
and distortions of the SNO day and night spectra. 
The neutron and low energy backgrounds~\cite{sno2} are included in $R_i^{th}$. 
 The $^8$B flux contribution to $R_i^{th}$ 
is multiplied by the normalization factor $\beta$ (relative to the central
value of the SSM $^8$B value), which is constrained
by the oscillation parameter dependent NC and CC flux components 
of the SNO day and night spectra. 

In Fig.~\ref{fig1}, we show the results of an analysis in which $\Delta m^2$,
$\tan^2 \theta$ and $\beta$ have been varied.
We only show those regions that are allowed at 3$\sigma$. Then 
the Small Mixing Angle (SMA) 
and VAC solutions are absent. 
We are left with only the LMA and LOW solutions. The
contours represent the 95.4\% C.L. (2$\sigma$), 99\% 
and 99.73\% C.L. (3$\sigma$)
allowed regions which correspond to 
$\Delta \chi^2= 6.17, 9.21$ and 11.83, respectively.
Values of $\theta$ larger than $\pi/4$ are not allowed at the 5$\sigma$ C.L. 
The best-fit parameter 
values from the analysis are presented in Table~\ref{tab2}.
It is significant that the LMA solution is favored over the LOW at the 99\%
C.L.
The correlation between the CC and NC rates extractable
from the SNO day and night spectra goes a long way in choosing the LMA over
the LOW. Table~\ref{tab2} shows that the data choose large 
NC/CC ratios in all MSW regions. An NC/CC ratio \mbox{$\sim$ 3} in the LOW
region is at the upper end of the range possible in this region.  
 
The 
best-fit, minimum and maximum values of the CC day-night asymmetry
{\footnote{The CC day-night asymmetry is defined by 
\begin{equation}
A_{DN}=2{N-D \over N+D}\,,
\end{equation}
where $D$ and $N$ are the total CC fluxes detected during the days and nights, respectively. Discussions of earth regeneration effects can
be found in Ref.~\cite{regeneration}.}, 
$A_{DN}(\beta)$, at 2$\sigma$ 
for the LMA region are displayed in Table~\ref{tab3}.

We next illustrate how well the best-fits of the two contending solutions, 
LMA and LOW, do in relation to the average survival probabilities of the 
high energy ($^8$B and $hep$), 
intermediate energy ($^7$Be, $pep$, $^{15}$O and $^{13}$N) and low 
energy ($pp$) neutrinos extracted from the experimental rates.
\begin{table}[ht]
\caption{Best-fit, minimum and maximum values of $A_{DN}(\beta)$ at 
2$\sigma$ for the LMA region.}
\begin{center}
\begin{tabular}{|l|c|c|c|c|} 
\hline 
 & $A_{DN}(\beta)$ &$\Delta m^2$ (eV$^2$) & $\tan^2 \theta$& $\beta$ \\ \hline
  at best-fit &0.043     &$5.6\times 10^{-5}$ & 0.39 & 1.09\\ \hline
   minimum at 2$\sigma$& 0. &$1.8\times 10^{-4}$ & 0.35 & 0.85\\ \hline
   maximum at 2$\sigma$&0.122  &$2.8\times 10^{-5}$ & 0.35 & 1.15 \\ \hline
\end{tabular}
\end{center}
\label{tab3}
\end{table}
 For a description of 
how these probabilities are obtained see Refs.~\cite{lma,piecing}. 
In Fig.~\ref{fig3}, we plot
each model-independently extracted survival probability at the mean energy 
 of the high, intermediate
and low energy neutrinos relevant to the experiments.
The vertical error bars result from the experimental 
uncertainties in the rate measurements and the theoretical 
uncertainties in the 
SSM flux predictions. The horizontal error bars span the energy ranges of
  high, intermediate and low energy neutrinos. The solid and dashed lines 
superimposed on the plot are flux-weighted survival probabilities 
at the SK detector corresponding to the best-fit LMA and LOW points 
of Table~\ref{tab2}, respectively. 
The monoenergetic $^7$Be and $pep$ fluxes
are not included in the averaging. The wiggles in the survival 
probabilities at high energies is a result of earth-matter effects. 
Aside from the averaging, a similar plot was
made in Ref.~\cite{berezinsky} with pre-SNO NC data. 

As a glimpse of the precision in the LMA region that
 KamLAND data may provide us with in three years, 
we have simulated data (with an antineutrino energy 
threshold of 3.3 MeV) at the 
best-fit LMA point and overlayed the expected 2$\sigma$, 99\% C.L.
 and 3$\sigma$ regions on the
currently allowed LMA region in Fig.~\ref{kamland}.

We conclude that the
LMA solution with nonmaximal mixing 
and a large active neutrino component in the solar flux is favored at the
99\% C.L.; 
the 2$\sigma$ allowed region spans 
\mbox{$2.7\times 10^{-5} - 1.8\times 10^{-4}$ eV$^2$} 
in $\Delta m^2$ and $0.27 - 0.55$ in $\tan^2 \theta$. KamLAND is an ideal
experiment to precisely measure oscillation parameters in this range.
No solution with $\theta \geq \pi/4$ 
is allowed at 5$\sigma$. Since the mixing is found to be nonmaximal, 
a determination of the neutrino mass scale via neutrinoless double
beta decay is an exciting possibility if neutrinos are Majorana~\cite{double}. 
A large sterile neutrino component
in the solar flux remains viable.  

\vspace{0.5in}
{\it Acknowledgements}. 
We thank M. Chen and J. Klein 
for several valuable discussions regarding SNO's global analysis.
We thank E. Kearns and M. Smy for providing us with the latest SK day and 
night spectra and for useful discussions. 
This research was supported by the U.S. Department of Energy
under Grants No.~DE-FG02-95ER40896, No.~DE-FG02-91ER40676 and
 No.~DE-FG02-01ER41155 and by the Wisconsin Alumni Research Foundation.

\newpage

\begin{figure}[t]
\centering\leavevmode
\mbox{\psfig{file=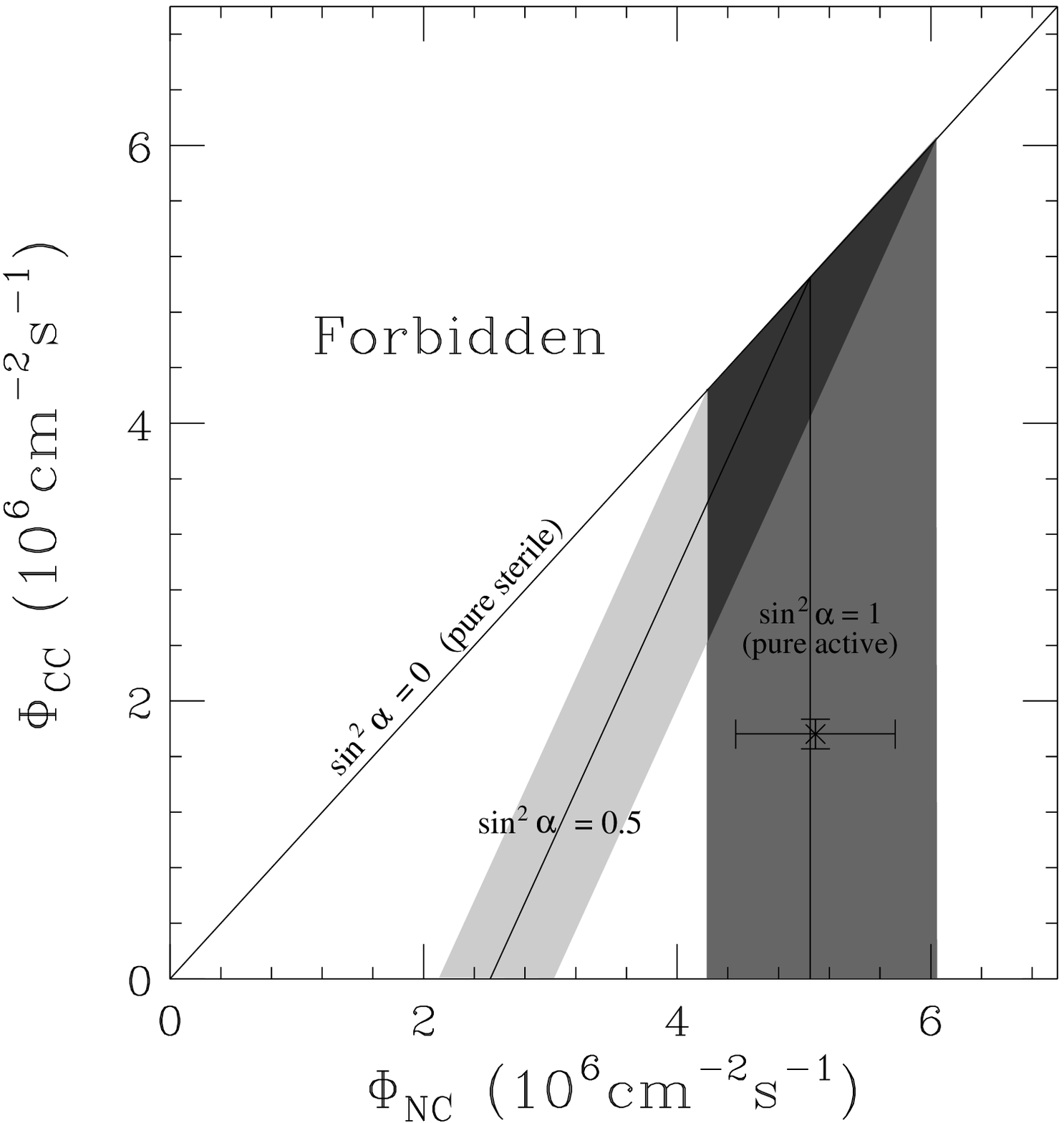,width=14cm,height=14cm}}
\medskip
\caption[]
{Graphical representation of Eq.~(\ref{ster}). The diagonal 
($\Phi_{NC}=\Phi_{CC}$) corresponds to 
$\sin^2 \alpha=0$ or pure sterile oscillations. The dark-shaded band
encloses the values of $\Phi_{NC}$ and $\Phi_{CC}$ allowed by the SSM at 1$\sigma$ 
for $\sin^2 \alpha=1$ or pure active oscillations. 
The light-shaded band is the region allowed by the SSM at 1$\sigma$ if $\sin^2 \alpha=0.5$. 
The SNO NC and CC measurements assuming an undistorted $^8$B spectrum 
(with 1$\sigma$ uncertainties) are marked with a cross. }
\label{sterile}
\end{figure}

\newpage
\begin{figure}[t]
\centering\leavevmode
\mbox{\psfig{file=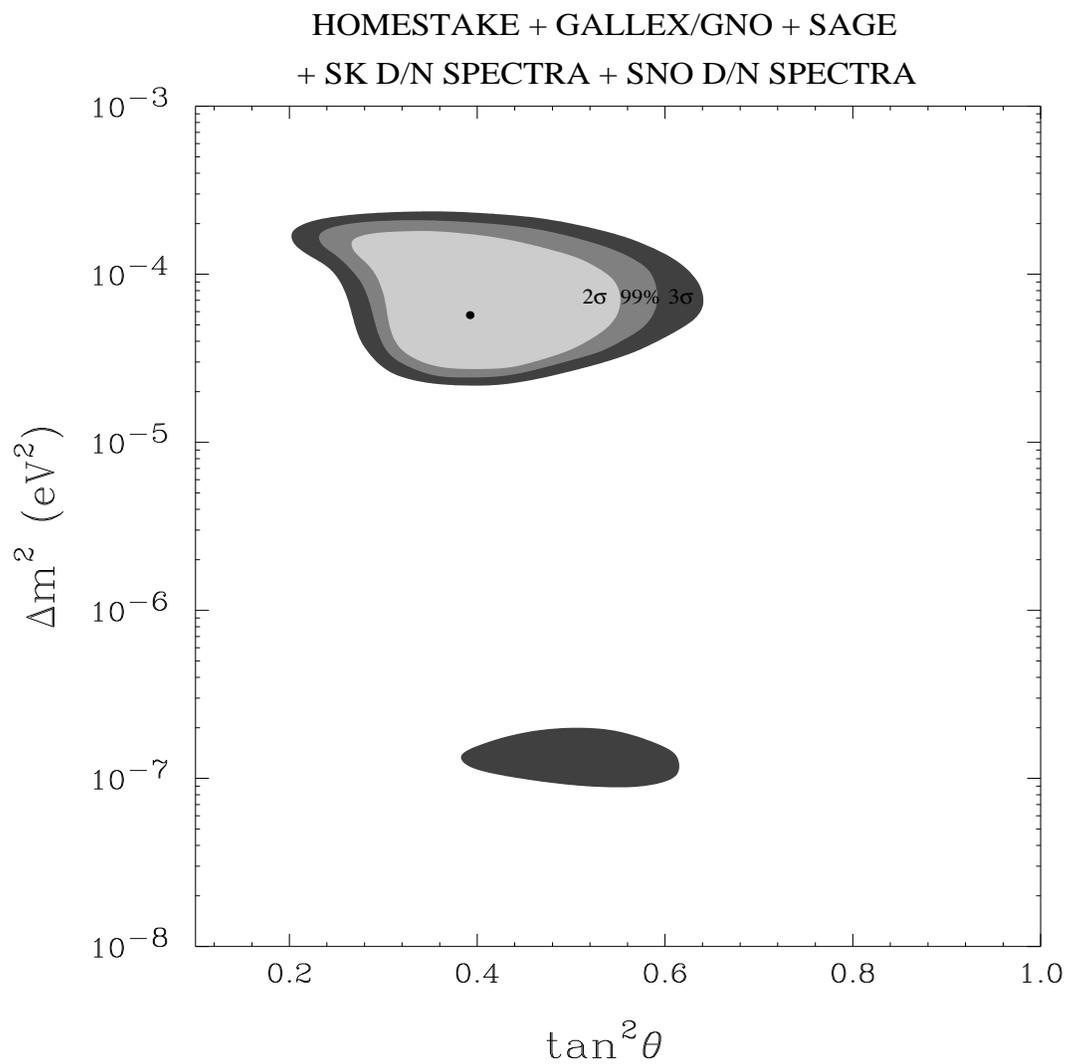,width=14cm,height=14cm}}
\medskip
\caption[]
{The 2$\sigma$,  99\% C.L. 
and 3$\sigma$ allowed regions from a fit to the Homestake, GALLEX+GNO
and SAGE rates, and the SK and SNO day and night spectra.}
\label{fig1}
\end{figure}

\begin{figure}[t]
\centering\leavevmode
\mbox{\psfig{file=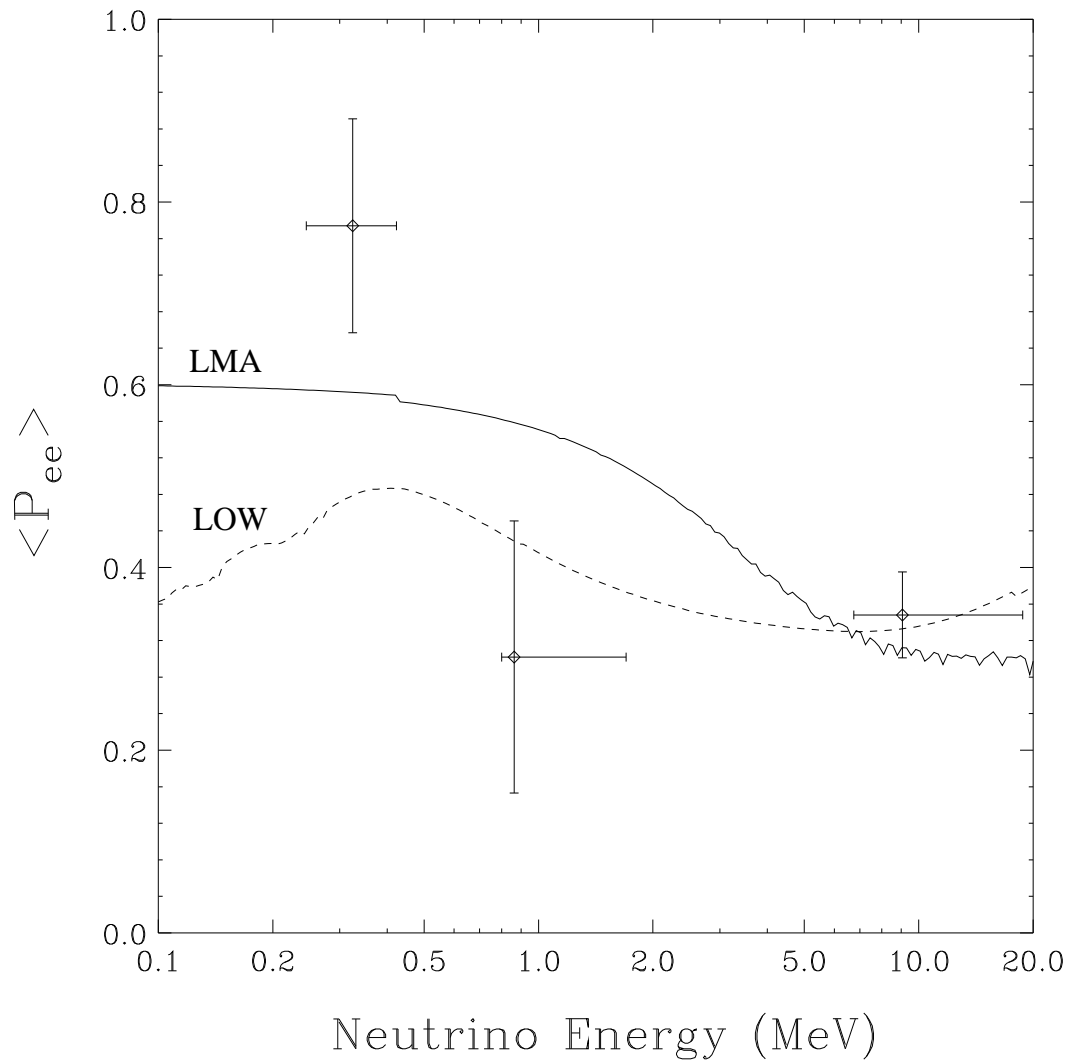,width=14cm,height=14cm}}
\medskip
\caption[]
{The flux-weighted 
survival probabilities of the best-fit LMA (solid) and LOW (dashed) 
solutions from Table~\ref{tab2} in relation to the model-independently extracted values
from the data.}
\label{fig3}
\end{figure}

\begin{figure}[t]
\centering\leavevmode
\mbox{\psfig{file=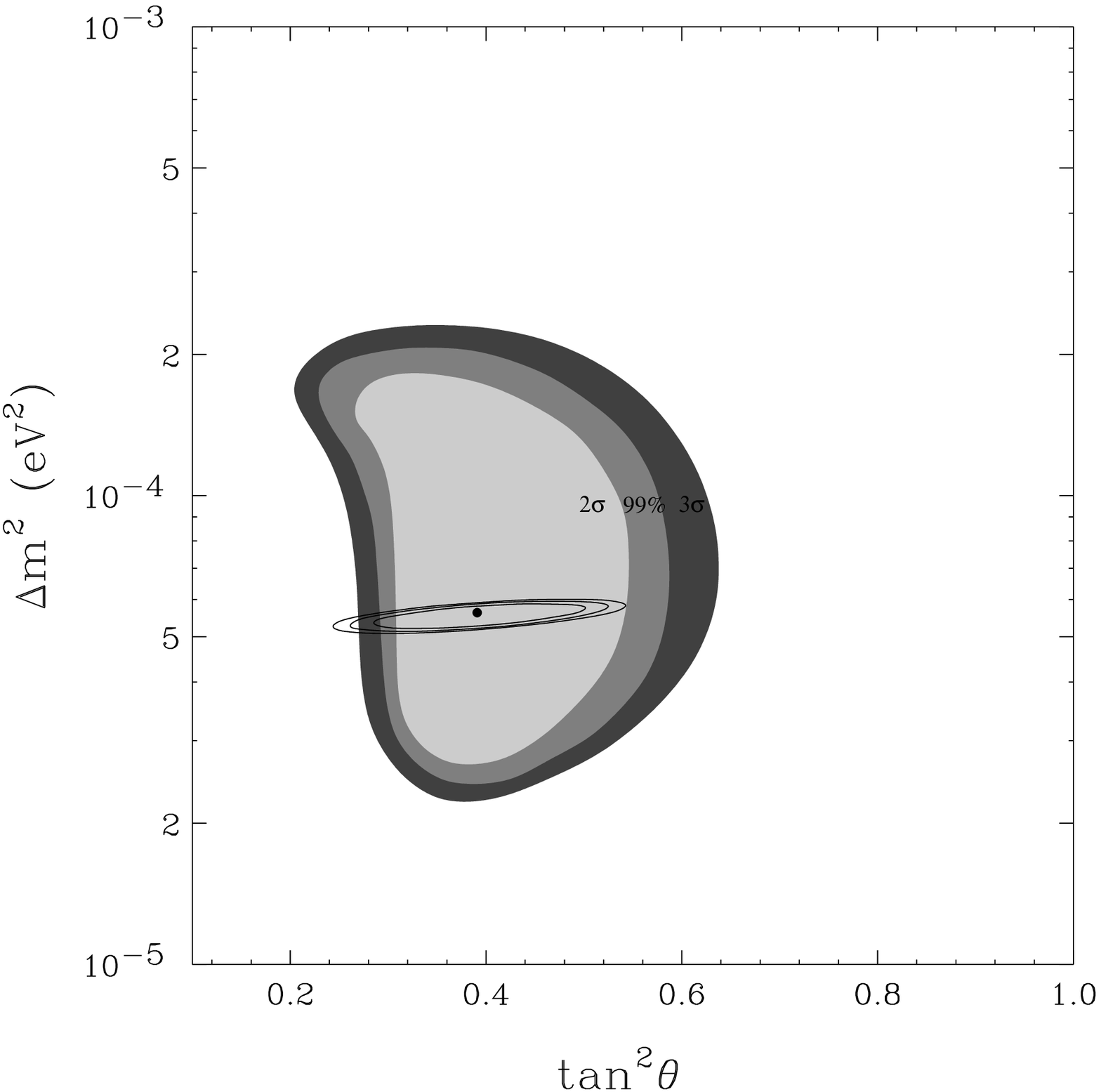,width=14cm,height=14cm}}
\medskip
\caption[]
{Projection of how well KamLAND will determine the oscillation 
parameters with three years of data accumulation assuming an LMA solution. 
Data were simulated at the best-fit LMA parameters. The ellipses are 
the 2$\sigma$, 99\% C.L. and 3$\sigma$ KamLAND regions.}
\label{kamland}
\end{figure}


\vspace{0.25in}
\newpage
\begin{thebibliography}{99}

\bibitem{SSM}
J.~N.~Bahcall, M.~H.~Pinsonneault and S.~Basu,
Astrophys.\ J.\  {\bf 555}, 990 (2001)
[arXiv:astro-ph/0010346];

\bibitem{cl}
B.~T.~Cleveland {\it et al.},
Astrophys.\ J.\  {\bf 496}, 505 (1998).

\bibitem{gallex}
W.~Hampel {\it et al.}  [GALLEX Collaboration],
Phys.\ Lett.\ B {\bf 447}, 127 (1999).

\bibitem{gno}
C. Cattadori, for the GNO collaboration, 
talk at TAUP 2001 conference, LNGS September 2001, to be
published in Nucl. Phys. B (Proc. Suppl.). 

\bibitem{sage}
J. N. Abdurashitov {\it et al.}  [SAGE Collaboration],
arXiv:astro-ph/0204245.

\bibitem{sk}
M.~B.~Smy,
arXiv:hep-ex/0202020;
S.~Fukuda {\it et al.}  [SuperKamiokande Collaboration],
Phys.\ Rev.\ Lett.\  {\bf 86}, 5651 (2001)
[arXiv:hep-ex/0103032].

\bibitem{sno}
Q.~R.~Ahmad {\it et al.}  [SNO Collaboration],
arXiv:nucl-ex/0204008.

\bibitem{sno1}
Q.~R.~Ahmad {\it et al.}  [SNO Collaboration],
arXiv:nucl-ex/0204009.

\bibitem{sno2}
SNO Collaboration; http://www.sno.phy.queensu.ca/sno/prlwebpage/

\bibitem{kamland}
The KamLAND proposal, Stanford-HEP-98-03.

\bibitem{bmw}
V.~Barger, D.~Marfatia and B.~P.~Wood,
Phys.\ Lett.\ B {\bf 498}, 53 (2001)
[arXiv:hep-ph/0011251].

\bibitem{chooz}
M.~Apollonio {\it et al.}  [CHOOZ Collaboration],
Phys.\ Lett.\ B {\bf 466}, 415 (1999)
[arXiv:hep-ex/9907037].

\bibitem{lma}
V.~Barger, D.~Marfatia and K.~Whisnant,
Phys.\ Rev.\ Lett.\  {\bf 88}, 011302 (2002)
[arXiv:hep-ph/0106207].

\bibitem{s17}
A.~R.~Junghans {\it et al.},
Phys.\ Rev.\ Lett.\  {\bf 88}, 041101 (2002)
[arXiv:nucl-ex/0111014].

\bibitem{robust}
J.~N.~Bahcall, M.~C.~Gonzalez-Garcia and C.~Pena-Garay,
arXiv:hep-ph/0111150.

\bibitem{int}
INT Mini-Wokshop: Neutrino Masses and Mixing; \\ 
http://int.phys.washington.edu/PROGRAMS/02-1mini.html

\bibitem{john}
J.~N.~Bahcall, M.~C.~Gonzalez-Garcia and C.~Pena-Garay,
arXiv:hep-ph/0204194.

\bibitem{err}
J.~N.~Bahcall, M.~C.~Gonzalez-Garcia and C.~Pena-Garay,
JHEP {\bf 0108}, 014 (2001)
[arXiv:hep-ph/0106258].

\bibitem{hep}
L.~E.~Marcucci, R.~Schiavilla, M.~Viviani, A.~Kievsky, S.~Rosati and J.~F.~Beacom,
Phys.\ Rev.\ C {\bf 63}, 015801 (2001)
[arXiv:nucl-th/0006005];
T.~S.~Park {\it et al.},
arXiv:nucl-th/0107012.

\bibitem{ind}
P.~I.~Krastev and A.~Y.~Smirnov,
arXiv:hep-ph/0108177;
J.~N.~Bahcall, P.~I.~Krastev and A.~Y.~Smirnov,
JHEP {\bf 0105}, 015 (2001)
[arXiv:hep-ph/0103179].

\bibitem{global}
In addition to Refs.~\cite{robust,err,ind}, see
G.L. Fogli, E. Lisi, D. Montanino and A. Palazzo, Phys.\ Rev.\ D {\bf 64},
093007 (2001) [arXiv:hep-ph/0106247];
P.~Aliani, V.~Antonelli, M.~Picariello and E.~Torrente-Lujan,
arXiv:hep-ph/0111418;
A.~M.~Gago, M.~M.~Guzzo, P.~C.~de Holanda, H.~Nunokawa, O.~L.~Peres, V.~Pleitez and R.~Zukanovich Funchal,
Phys.\ Rev.\ D {\bf 65}, 073012 (2002)
[arXiv:hep-ph/0112060];
A.~Bandyopadhyay, S.~Choubey, S.~Goswami and D.~P.~Roy,
arXiv:hep-ph/0203169.

\bibitem{analytic}
S.~J.~Parke,
Phys.\ Rev.\ Lett.\  {\bf 57}, 1275 (1986);
W.~C.~Haxton, Phys.\ Rev.\ Lett.\  {\bf 57}, 1271 (1986);
Phys.\ Rev.\ D {\bf 35}, 2352 (1987); 
S.~Toshev,
Phys.\ Lett.\ B {\bf 196}, 170 (1987);
S.~T.~Petcov,
Phys.\ Lett.\ B {\bf 200}, 373 (1988).

\bibitem{prem}
A. Dziewonski and D. Anderson, Phys. Earth Planet. Inter. {\bf 25}, 297 (1981).

\bibitem{exp}
J.~N.~Bahcall and P.~I.~Krastev,
Phys.\ Rev.\ C {\bf 56}, 2839 (1997)
[arXiv:hep-ph/9706239].

\bibitem{B8}
C.~E.~Ortiz, A.~Garcia, R.~A.~Waltz, M.~Bhattacharya and A.~K.~Komives,
Phys.\ Rev.\ Lett.\  {\bf 85}, 2909 (2000)
[arXiv:nucl-ex/0003006].

\bibitem{es}
J.~N.~Bahcall, M.~Kamionkowski and A.~Sirlin,
Phys.\ Rev.\ D {\bf 51}, 6146 (1995)
[arXiv:astro-ph/9502003].

\bibitem{cs}
S.~Nakamura, T.~Sato, S.~Ando, T.~S.~Park, F.~Myhrer, V.~Gudkov and K.~Kubodera,
arXiv:nucl-th/0201062.

\bibitem{respsk}
M.~Nakahata {\it et al.}  [Super-Kamiokande Collaboration],
Nucl.\ Instrum.\ Meth.\ A {\bf 421}, 113 (1999)
[arXiv:hep-ex/9807027].

\bibitem{corr}
G.~L.~Fogli and E.~Lisi,
Astropart.\ Phys.\  {\bf 3}, 185 (1995).

\bibitem{regeneration}
A.~J.~Baltz and J.~Weneser,
Phys.\ Rev.\ D {\bf 35}, 528 (1987);
Phys.\ Rev.\ D {\bf 37}, 3364 (1988);
J.~N.~Bahcall, P.~I.~Krastev and A.~Y.~Smirnov,
Phys.\ Rev.\ D {\bf 62}, 093004 (2000)
[arXiv:hep-ph/0002293];
M.~Maris and S.~T.~Petcov,
Phys.\ Rev.\ D {\bf 62}, 093006 (2000)
[arXiv:hep-ph/0003301];
V.~Barger, D.~Marfatia, K.~Whisnant and B.~P.~Wood,
Phys.\ Rev.\ D {\bf 64}, 073009 (2001)
[arXiv:hep-ph/0104095].

\bibitem{piecing}
V.~Barger, D.~Marfatia and K.~Whisnant,
Phys.\ Lett.\ B {\bf 509}, 19 (2001)
[arXiv:hep-ph/0104166].

\bibitem{berezinsky}
V.~Berezinsky and M.~Lissia,
Phys.\ Lett.\ B {\bf 521}, 287 (2001)
[arXiv:hep-ph/0108108].

\bibitem{double}
V.~Barger, S.~L.~Glashow, D.~Marfatia and K.~Whisnant,
arXiv:hep-ph/0201262.

\end{thebibliography}
\end{document}